\documentclass[twocolumn,english]{revtex4-1}
\usepackage[T1]{fontenc}
\usepackage[latin9]{inputenc}
\usepackage{verbatim}
\usepackage{float}
\usepackage{amsmath}
\usepackage{amssymb}
\usepackage{graphicx}
\usepackage{esint}

\makeatletter

\providecommand{\tabularnewline}{\\}
\floatstyle{ruled}
\newfloat{algorithm}{H}{loa}
\providecommand{\algorithmname}{Algorithm}
\floatname{algorithm}{\protect\algorithmname}

\usepackage[caption=false]{subfig}
\usepackage{babel}

\@ifundefined{showcaptionsetup}{}{%
 \PassOptionsToPackage{caption=false}{subfig}}
\usepackage{subfig}
\makeatother

\usepackage{babel}
\begin{document}
\title{A polynomial time algorithm for studying physical observables in chaotic
eigenstates}
\author{Pavan Hosur}
\affiliation{Department of Physics, University of Houston, Houston, TX 77204}
\affiliation{Texas Center for Superconductivity, Houston, TX 77204}
\date{\today}
\begin{abstract}
We introduce an algorithm, the Orthogonal Operator Polynomial Expansion
(OOPEX), to approximately compute expectation values in energy eigenstates
at finite energy density of non-integrable quantum many-body systems
with polynomial effort, whereas exact diagonalization (ED) of the
Hamiltonian $H$ is exponentially hard. The OOPEX relies on the eigenstate
thermalization hypothesis, which conjectures that eigenstate expectation
values of physical observables in such systems vary smoothly with
the eigenstate energy (and other macroscopic conserved quantities,
if any), and computes them through a series generated by repeated
multiplications, rather than diagonalization, of $H$ and whose successive
terms oscillate faster with the energy. The hypothesis guarantees
that only the first few terms of this series contribute appreciably.
We further show that the OOPEX, in a sense, is the most optimum algorithm
based on series expansions of $H$ as it avoids computing the many-body
density of states which plagues other similar algorithms. Then, we
argue non-rigorously that working in the Fock space of operators,
rather than that of states as is usually done, yields convergent results
with computational resources that scale polynomially with $N$. We
demonstrate the polynomial scaling by applying the OOPEX to the non-integrable
Ising chain and comparing with ED and high-temperature expansion (HTX)
results. The OOPEX provides access to much larger $N$ than ED and
HTX do, which facilitates overcoming finite-size effects that plague
the other methods to extract correlation lengths in chaotic eigenstates.
In addition, access to large systems allows testing a recent conjecture
that the Renyi entropy of chaotic eigenstates has positive curvature
if the Renyi index $>1$, and we find encouraging supporting evidence.
\end{abstract}
\maketitle

\section{Introduction}

Some quantum many-body systems are integrable, i.e., they contain
simplifying properties such as easy-to-diagonalize conserved operators,
emergent conservation laws resulting from strong disorder \citep{MBLParameswaranVasseur,MBLreviewHuseNandkishore,Pal10,ALET2018}
or factorizable scattering matrices \citep{Bethe1931,GUTKIN1987,LeeGuandelCampo2012,Sutherland1995}
that make them computationally -- and sometimes analytically --
tractable. Most lack these properties and are said to be non-integrable
(NI). Recent years have revealed that energy eigenstates with finite
energy density in NI quantum many-body systems provide portals into
diverse areas of physics and related fields. For instance, their properties
relevant to condensed matter, quantum information, fundamental physics,
gravity and statistical mechanics, respectively, include the facts
that they encode finite temperature phase transitions \citep{Fratus2015,Fratus2016},
form a quantum error correcting code \citep{Pastawski:2015qua,Almheiri:2014lwa,Brandao2019,Bao:2019aa},
enable reconstruction of the entire Hamiltonian \citep{Garrison2018,Qi2019determininglocal},
mimic conformal field theories \citep{Lashkari2018,Hikida2018,Datta2019,Lashkari_2018}
which in turn mimic quantum gravity under the holographic mapping
\citep{Maldacena1999,QiExactHolographic} and resemble equilibrium
statistical ensembles if only simple measurements are made \citep{Popescu2006,Linden2009,Deutsch1991,Deutsch2010,Neumann1929,QETproof,vonNeumann2010,DAlessioReview,Reimann2007,Reimann2015},
where ``simple'' usually means few-body and local, and includes
observables that real experiments can measure. Such eigenstates are
also relevant to chaos, which earns then the name chaotic eigenstates.
Firstly, if a quantum system has a well-defined classical limit and
the classical system is chaotic, the quantum eigenstates are expected
to satisfy the eigenstate thermalization hypothesis (ETH) \citep{Srednicki1994,Srednicki1996,Srednicki1999}.
Secondly, quantum systems with ETH-satisfying eigenstates exhibit,
in many cases, temporal correlations that resemble the famous ``butterfly
effect'' from classical chaos \citep{MaldacenaStanfordSYK,Polchinski2016,Roberts2016,Roberts:2014ifa,Shenker:2014aa,Xu2019,Foini2019,DAlessioReview,Maldacena:2015waa,Hosur2016a}.
These unique properties make simulating chaotic eigenstates an important
goal of quantum many-body physics.

Unfortunately, this is a Herculean task. Chaotic eigenstates occur
at finite energy density above the ground state, which puts them beyond
the reach of the numerous powerful algorithms available for studying
ground state and low-energy physics. Quantum Monte Carlo methods \emph{can}
study physics at finite energy density with polynomial effort in $N$,
the number of degrees of freedom, if a suitable discrete symmetry
cures the sign-problem \citep{Troyer2005,SignProblem}. If there is
no symmetry -- in which case the model is maximally NI -- the sign-problem
persists and the complexity becomes exponential. Finally, the lack
of simplifying properties in NI systems makes brute force exact diagonalization
(ED) of $H$ exponentially hard. Thus, the problem of simulating chaotic
eigenstates is generally deemed unsolvable.

In this work, we introduce an algorithm -- the Orthogonal Operator
Polynomial Expansion (OOPEX) -- that extracts useful information
from chaotic eigenstates with polynomial effort. It achieves this
efficiency by exploiting the ETH, which states that $\left\langle A(E_{i})\right\rangle $,
the expectation value of any simple operator $A$ in an energy eigenstate
$|E_{i}\rangle$ of a NI Hamiltonian $H$, acquires the same value
in nearby eigenstates at finite energy density in the thermodynamic
limit: 
\begin{equation}
\left\langle A(E_{i})\right\rangle \xrightarrow{N\to\infty}\left\langle A(E_{j})\right\rangle \text{ if }\frac{E_{i}-E_{0}}{N}\xrightarrow{N\to\infty}\frac{E_{j}-E_{0}}{N}\neq0\label{eq:ETH}
\end{equation}
where $E_{0}$ is the ground state energy \citep{Srednicki1999,Deutsch1991,Reimann2015,Garrison2018}.
For systems with a bounded spectrum such as lattice models, (\ref{eq:ETH})
is expected when $E_{0}$ refers to the highest energy state as well.
Specifically, we will express $\rho(E_{i})=\left|E_{i}\left\rangle \right\langle E_{i}\right|$
as a power series in $H$, modified such that higher order terms capture
progressively more complicated observables. As a result, truncating
the series retains only the simple, ETH-satisfying, experimentally
accessible observables. In contrast, ED computes the full wavefunction
exactly before extracting simple observables from it. This unnecessary
computation is the source of ED's inefficiency. ED also requires storing
$H$ as a matrix in the local Fock basis, which consumes an exponential
amount of memory. Here, we use the Operator Fock Space Representation
(OFSR) \citep{Hosur2016,Ros2015}, which eliminates the need to store
and manipulate state-vectors or operator-matrices and consequently
reduces computational needs to merely polynomial in $N$. Crucially,
we show that the OFSR is the natural language for developing the OOPEX.

The OOPEX is distinct from a simple high-temperature expansion (HTX),
which entails Taylor expanding $e^{-\beta H}$ in powers of $\beta$
but fails to exploit non-integrability of the system. As explained
in Sec. \ref{subsec:Optimum-polynomial-expansion}, it also differs
crucially from other polynomial expansion methods by avoiding computing
the density of states $D(E)$. $D(E)$ is usually not well-approximated
by polynomials, but algorithms such as the kernel polynomial method
\citet{WeissKPMReview} find a polynomial approximant to it nonetheless
and thereby converge much slower.

\section{The Algorithm}

\subsection{ETH-based truncation}

Suppose our goal is to compute $\left\langle A(E_{i})\right\rangle =\text{tr}[\rho(E_{i})A]$.
If the spectrum of $H$ lacks degeneracies, as is expected for NI
systems, the Krylov space defined by $1,H,H^{2}\dots H^{d-1}$, where
$d$ is the total Hilbert space dimension, forms a complete basis
for the space of operators that commute with $H$. An alternate basis
for this space is simply $\rho(E_{i}),i=1\dots d$. Therefore, $\rho(E_{i})$
is expressible as a power series in $H$. A simple power series, however,
does not produce progressively diminishing contributions to $\left\langle A(E_{i})\right\rangle $,
so its truncation error is uncontrolled. To rectify this problem,
we first orthonormalize the Krylov space and write 
\begin{equation}
\rho(E_{i})=\sum_{m=0}^{d-1}p_{m}(E_{i})p_{m}(H)\label{eq:rho-series}
\end{equation}
where $p_{m}(x)=\sum_{k=0}^{m}a_{km}x^{k}$ is an $m^{th}$ degree
polynomial of its argument that satisfies the orthogonality conditions:
$\text{tr}\left[p_{m}(H)p_{m'}(H)\right]=\delta_{mm'}$, $\sum_{m=0}^{d-1}p_{m}(E_{i})p_{m}(E_{j})=\delta_{ij}$.
Intuitively, $i$ and $m$ are conjugate variables with respect to
the definition (\ref{eq:rho-series}), analogous to the conjugacies
of frequency and time with respect to Fourier transformation. While
exact, (\ref{eq:rho-series}) is impractical because $d$ grows exponentially
with $N$. We now argue, and later demonstrate using the NI Ising
model, that $O(1)$ terms suffice in practice. Then, (\ref{eq:rho-series})
involves computing only the first few powers of $H$ via multiplication,
which is far more efficient than diagonalizing it.

\begin{figure}
\begin{centering}
\includegraphics[width=0.9\columnwidth]{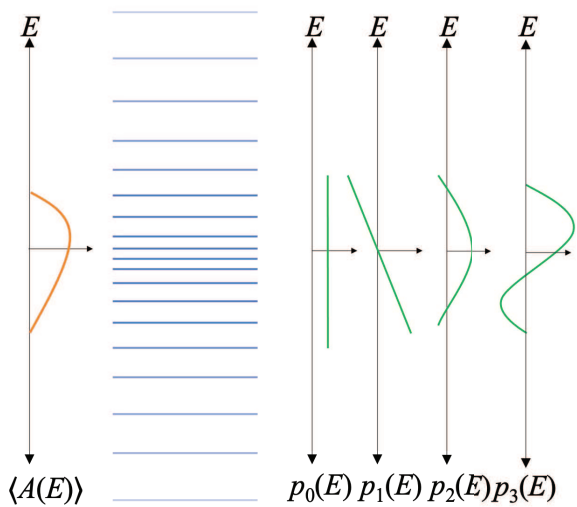} 
\par\end{centering}
\caption{Schematic of the OOPEX philosophy. $\left\langle A(E)\right\rangle $
varies slowly with $E$ far from the edges of the spectrum if $A$
satisfies the ETH. $p_{m}(E)$ are polynomials with $m$ roots, so
they oscillate faster with $E$ as $m$ increases. Thus, according
to (\ref{eq:rho-series}), $\left\langle A(E)\right\rangle $ receives
dominant contributions from small $m$.\label{fig:OOPEX Schematic}}
\end{figure}

To see why only the first few terms suffice, recall that $p_{m}(E_{i})$
is a polynomial in $E_{i}$ of degree $m$, so it varies slowly (rapidly)
with $E_{i}$ for small (large) $m$. Alternately, in analogy with
Fourier transformation, $p_{m}(E)$ with small $m$ has smooth $E$-dependence
whereas $p_{m}(E)$ with large $m$ will oscillate rapidly with $E$.
Therefore, if $\left\langle A(E)\right\rangle $ varies smoothly with
$E$ over a small energy window $\epsilon$, it will receive contributions
mainly from the first few terms in (\ref{eq:rho-series}). This will
allow us to truncate (\ref{eq:rho-series}) and make the OOPEX a viable
method. The philosophy is depicted in Fig. \ref{fig:OOPEX Schematic}.
Physically, the truncation discards information that distinguishes
between nearby eigenstates, but this information is stored in complicated
observables that are impossible to measure in practice anyway \citep{Hosur2016}.

How many terms must we retain without incurring significant truncation
error? We can crudely estimate an upper bound on $m_{c}$, the value
of $m$ at which convergence occurs, as follows. Energy is extensive,
$E\propto N$, while $\left\langle A(E)\right\rangle $ varies negligibly
over any sub-extensive interval $\epsilon\propto N^{\alpha};\alpha\to1^{-}$
according to (\ref{eq:ETH}). Crudely assuming that the $m$ roots
of $p_{m}(E)$ are real and equally spaced across the spectrum, $\epsilon$
will contain a root if $m>E/\epsilon\propto N^{1-\alpha}\ll N$. Choosing
$m_{c}\sim N^{1-\alpha}=O(1)$ as $\alpha\to1^{-}$, $\left\langle A(E)\right\rangle $
will receive both positive and negative contributions from the interval
$\epsilon$ for $m>m_{c}$. The net contribution will thus be small,
signaling convergence.

This crude estimate receives two competing refinements in practice:
(i) the cancellation of positive and negative contributions to $\left\langle A(E)\right\rangle $
within the window $\epsilon$ for $m>E/\epsilon$ is not exact, which
means more terms must be retained in (\ref{eq:rho-series}) to achieve
convergence; and (ii) the roots of $p_{m}(E)$ cluster near the middle
of the spectrum, which means some cancellation occurs even when $m<E/\epsilon$,
thereby decreasing $m_{c}$. In Sec. \ref{sec:Ising-model-results},
we find numerically for the NI Ising model that $m_{c}\lesssim3=O(1)$
indeed.

\subsection{Compression using OFSR}

So far, we have reduced the computation from diagonalization of $H$
to repeated multiplications of $H$, but the runtime and storage costs
are still exponential because $H$, written as a sparse matrix in
a local basis, has at least $O(d)$ terms. To reduce these costs,
we work in the OFSR \citep{Hosur2016,Ros2015}, in which operators
are expressed as vectors in operator Hilbert space: 
\begin{equation}
A=\sum_{\ell}\alpha_{\ell}\mathcal{O}_{\ell}\to||A\rangle\rangle=(\alpha_{1},\alpha_{2}\dots)^{T}
\end{equation}
where each $\mathcal{O}_{\ell}$ is a product of local operators,
$\text{tr}\left(\mathcal{O}_{\ell}^{\dagger}\mathcal{O}_{\ell'}\right)=\delta_{\ell\ell'}$
and the notation $||\dots\rangle\rangle$ has been defined to denote
vectors in operator Hilbert space. For instance, basis operators for
an $N$-site lattice with spin-1/2 on each site can be taken to be
$\mathcal{O}_{\ell}=2^{-N/2}\prod_{i\in\text{sites}}^{\otimes}\sigma_{i}^{\alpha}$,
where $\sigma_{i}^{\alpha}$ is either a $2\times2$ identity matrix
or a Pauli matrix, and $\prod^{\otimes}$ denotes an outer product.
The OFSRs of a basis operator, a local Hamiltonian and a Hamiltonian
with long-range $p$-body interactions contain a single term, $O(N)$
terms and $O(N^{p})$ terms, respectively, as opposed to $O(d)$ terms
in their usual matrix representation in a local basis. As a result,
the OFSR reduces storage costs from $O(d)$ to $O(N^{pm_{c}})$ if
we truncate (\ref{eq:rho-series}) at $m_{c}$, which is polynomial
in $N$ for $m_{c}=O(1)$. Naturally, the runtime is polynomial too
since only polynomially large vectors are manipulated.

The OFSR is the natural language for developing the OOPEX, because
each step of the algorithm, summarized in Algorithm \ref{alg:Main-steps},
has a simple interpretation in terms of the linear algebra of the
OFSR-vectors. For instance, applying standard $QR$-decomposition
on the matrix $\left(||1\rangle\rangle,||H\rangle\rangle,||H^{2}\rangle\rangle,\dots\right)$
yields the orthonormalized Krylov space $Q=\left(||1\rangle\rangle,||p_{1}(H)\rangle\rangle,||p_{2}(H)\rangle\rangle\dots\right)$
as well as the coefficients $a_{km}=\left(R^{-1}\right)_{km}$ and
hence, the polynomials $p_{m}(E)$. Moreover, the trace of a product
of operators reduces to the inner product of their OFSRs: $\text{tr}\left(B^{\dagger}A\right)\equiv\left\langle \left\langle B||A\right\rangle \right\rangle $,
which allows computing $\left\langle A(E)\right\rangle $ easily by
choosing $B=\rho(E)$. The main trade-off is that the rules for multiplying
operators in their OFSRs must be derived from the non-commutative
algebra of the basis operators. We find that this added cost is easily
overcome by the other gains. In contrast, the OFSR is not useful for
diagonalization-based algorithms such as ED because diagonalization
of a matrix does not correspond to any obvious operation on its OFSR-vector.
\begin{algorithm}
\begin{enumerate}
\item Express $H$ as a column vector in its OFSR, $||H\rangle\rangle$. 
\item Compute the Krylov space $K=\left\{ ||1\rangle\rangle,||H\rangle\rangle,\dots,||H^{m_{max}}\rangle\rangle\right\} $
for pre-selected $m_{max}$ via repeated multiplication with $|H\rangle$.
The multiplication rules are determined by the algebra of the OFSR
basis operators. 
\item Decompose $K$ as $K=QR$ where $Q$ is a $4^{N}\times(m_{max}+1)$
orthogonal matrix and $R$ is a $(m_{max}+1)\times(m_{max}+1)$ upper-triangular
matrix.
\begin{enumerate}
\item $Q$ is precisely the orthonormalized Krylov space: $Q=\left(||p_{0}(H)\rangle\rangle,||p_{1}(H)\rangle\rangle,\dots,||p_{m_{max}}(H)\rangle\rangle\right)$. 
\item $R$ provides $p_{m}(E)$ as $p_{m}(E)=\sum_{k=0}^{m}\left(R^{-1}\right)_{km}E^{k}$.
\item Using $Q$ and $p_{m}(E)$, determine $||\rho(E)\rangle\rangle$ using
(\ref{eq:rho-series}). 
\end{enumerate}
\item Compute the inner product $\langle\langle\rho(E)||A\rangle\rangle=\text{tr}[\rho(E)A]$. 
\end{enumerate}
\caption{Main steps of the OOPEX algorithm.\label{alg:Main-steps}}
\end{algorithm}

\subsection{Optimum polynomial expansion\label{subsec:Optimum-polynomial-expansion}}

In this section, we place the OOPEX in the broad context of polynomial
expansion methods. We show that the OOPEX, unlike other methods, avoids
computing the density of states. This eliminates a major source of
error and is presumably responsible for rapid convergence.

Consider expressing $A(E)=\left\langle E|A|E\right\rangle $ in terms
of pre-selected functions $q_{m}(E)$ that are orthogonal with respect
to the weight $w(E)$ over an interval $E\in[-E_{0},E_{0}]$. One
can always shift and rescale the Hamiltonian $H$ so that all the
energies $E_{i}$ lie in this interval. $A(E)$ can be written as
\begin{equation}
A(E)=\frac{w(E)}{D(E)}\sum_{m}\mu_{m}^{A}q_{m}(E)\label{eq:A(E)}
\end{equation}
where
\begin{equation}
\intop_{-E_{0}}^{E_{0}}q_{m}(E)q_{m'}(E)w(E)dE=N_{m}\delta_{mm'}
\end{equation}
and $D(E)=\sum_{i}\delta(E-E_{i})$ is the density of states. The
unknowns above are the moments $\mu_{m}^{A}$ and $D(E)$. $\mu_{m}^{A}$
are given by
\begin{equation}
\mu_{m}^{A}=\frac{1}{N_{m}}\intop_{-E_{0}}^{E_{0}}A(E)q_{m}(E)D(E)dE=\frac{1}{N_{m}}\text{tr}\left[Aq_{m}(H)\right]\label{eq:mu_m}
\end{equation}
The last expression is relatively easy to compute since it is simply
the Hilbert-Schmidt inner product of $A$ and $q_{m}(H)$, or the
inner product $\left\langle \left\langle A||q_{m}(H)\right\rangle \right\rangle $
in terms of their OFSRs. To determine $D(E)$, one chooses $A=1$,
which gives 
\begin{equation}
D(E)=w(E)\sum_{m}\mu_{m}^{D}q_{m}(E)\implies\mu_{m}^{D}=\frac{1}{N_{m}}\text{tr}\left[q_{m}(H)\right]
\end{equation}
Thus, computing $A(E)$ entails separately computing the moments $\mu_{m}^{A}$
and $\mu_{m}^{D}$, using the latter to determine $D(E)$, and finally
using (\ref{eq:A(E)}).

The OOPEX simplifies the above process by effectively choosing $w(E)=D(E)$
and $N_{m}=1$. Then, 
\begin{equation}
A(E)=\sum_{m}\text{tr}[Aq_{m}(H)]q_{m}(E)
\end{equation}
where
\begin{equation}
\intop_{-E_{0}}^{E_{0}}q_{m}(E)q_{m'}(E)D(E)dE=\text{tr}[q_{m}(H)q_{m'}(H)]=\delta_{mm'}\label{eq:qm-ortho}
\end{equation}
In other words, $q_{m}(E)$ defined by (\ref{eq:qm-ortho}) are precisely
the $p_{m}(E)$ defined in (\ref{eq:rho-series}). Note that the OOPEX
never explicitly calculates $D(E)$. Thus, it avoids a major source
of error compared to other methods that approximate $D(E)$ and achieve
faster convergence.

A well-known example of such an algorithm is the kernel polynomial
method which rescales energy so that $E_{0}=1$ and uses $q_{m}(E)=T_{m}(E)/\sqrt{1-E^{2}}$,
where $T_{m}(E)$ is the $m^{th}$ Chebyshev polynomial of the first
kind. $q_{m}(E)$ are orthonormal with respect to the weight $w(E)=\pi\sqrt{1-E^{2}}$
and are normalized as $N_{m}=(1+\delta_{m,0})/2$. If $H$ is an infinite
dimensional random hermitian matrix, then Wigner's semicircle law
states that $D(E)=\frac{2}{\pi}\sqrt{1-E^{2}}$, so that $w(E)/D(E)=\pi^{2}/2$.
Then, the kernel polynomial method is equivalent to the OOPEX up to
an overall factor of $\pi^{2}/2$ that can be absorbed into $N_{m}$.
However, $D(E)$ differs significantly from the semi-circle law for
realistic systems with the local Hamiltonian, and accurately computing
$D(E)$ can require hundreds of Chebyshev moments $\mu_{m}^{D}$.
In contrast, the OOPEX requires calculating only the moments $\mu_{m}^{A}$
since $\mu_{m}^{D}=\frac{1}{N_{m}}\text{tr}[q_{m}(H)]=\delta_{m,0}$
is trivially known.

\section{Ising model results\label{sec:Ising-model-results}}

We now demonstrate the OOPEX on a prototypical NI spin model, namely,
the 1D Ising model with transverse and longitudinal fields, given
by 
\begin{equation}
H=\sum_{r}\left(J\sigma_{r}^{z}\sigma_{r+1}^{z}+h_{x}\sigma_{r}^{x}+h_{z}\sigma_{r}^{z}\right)\label{eq:Hamiltonian}
\end{equation}
where $\{\sigma_{r}^{\alpha}\}$ are Pauli matrices. $H$ is integrable
if any one of $J$, $h_{x}$ and $h_{z}$ vanishes, but is NI otherwise.
This model is ideal for demonstrating the OOPEX because it does not
harbor any non-analyticities such as phase transitions at finite temperatures.
As a result, the analytic expansion in (\ref{eq:rho-series}) is expected
to converge quickly. We choose $J=1$, $h_{x}=-1.05$ and $h_{z}=0.5$,
and open boundary conditions to prevent momentum conservation. Then,
$||H\rangle\rangle$ contains $3N-1$ non-zero terms with $N$, $N$
and $N-1$ terms equal to $h_{x}$, $h_{z}$ and $J$, respectively
and can therefore be stored as a sparse vector of length $4^{N}$
with only $3N-1$ non-zero elements. Computing $||H^{m}\rangle\rangle$
entails evaluating indices while keeping track of the non-commutative
algebra of the Pauli operators, which is a main computational cost,
but requires storing only $\sim(3N)^{m}$ real numbers. Table \ref{tab:OFSR-of-H}
shows the explicit OFSRs of $H$ and $H^{2}$ for $N=2$. All calculations
were performed on a 2.7 GHz 12-core processor with 64 GB random access
memory.
\begin{table}
\begin{centering}
\begin{tabular}{|c|c|c|c|}
\hline 
Operator & Index & $||H\rangle\rangle$ & $||H^{2}\rangle\rangle$\tabularnewline
\hline 
\hline 
$\mathbb{1}$ & 00 & - & $J^{2}+2h_{x}^{2}+2h_{z}^{2}$\tabularnewline
\hline 
$\sigma_{2}^{x}$ & 01 & $h_{x}$ & -\tabularnewline
\hline 
$\sigma_{2}^{y}$ & 02 & - & -\tabularnewline
\hline 
$\sigma_{2}^{z}$ & 03 & $h_{z}$ & $2Jh_{z}$\tabularnewline
\hline 
$\sigma_{1}^{x}$ & 10 & $h_{x}$ & -\tabularnewline
\hline 
$\sigma_{1}^{x}\sigma_{2}^{x}$ & 11 & - & $2h_{x}^{2}$\tabularnewline
\hline 
$\sigma_{1}^{x}\sigma_{2}^{y}$ & 12 & - & -\tabularnewline
\hline 
$\sigma_{1}^{x}\sigma_{2}^{z}$ & 13 & - & $2h_{x}h_{z}$\tabularnewline
\hline 
$\sigma_{1}^{y}$ & 20 & - & -\tabularnewline
\hline 
$\sigma_{1}^{y}\sigma_{2}^{x}$ & 21 & - & -\tabularnewline
\hline 
$\sigma_{1}^{y}\sigma_{2}^{y}$ & 22 & - & -\tabularnewline
\hline 
$\sigma_{1}^{y}\sigma_{2}^{z}$ & 23 & - & -\tabularnewline
\hline 
$\sigma_{1}^{z}$ & 30 & $h_{z}$ & $2Jh_{z}$\tabularnewline
\hline 
$\sigma_{1}^{z}\sigma_{2}^{x}$ & 31 & - & $2h_{x}h_{z}$\tabularnewline
\hline 
$\sigma_{1}^{z}\sigma_{2}^{y}$ & 32 & - & -\tabularnewline
\hline 
$\sigma_{1}^{z}\sigma_{2}^{z}$ & 33 & $J$ & $2h_{z}^{2}$\tabularnewline
\hline 
\end{tabular}
\par\end{centering}
\caption{OFSR of $H$, given by (\ref{eq:Hamiltonian}), and $H^{2}$ for $N=2$.
The operators are mapped to $N$-digit base-4 integers as $\sigma_{r}^{0}\to0_{r}$,
$\sigma_{r}^{x}\to1_{r}$, $\sigma_{r}^{y}\to2_{r}$, $\sigma_{r}^{z}\to3_{r}$,
while the coefficients become the non-zero entries in a sparse vector
indexed by the base-4 integers. Even though the overall length of
the sparse vector grows as $4^{N}$, the OOPEX only involves $||H^{m}\rangle\rangle$
for small $m$, so the number of non-zero entries that need to be
stored grows mildly with $N$ as $\sim(3N)^{m}$.\label{tab:OFSR-of-H}}
\end{table}

\subsection{Observables}

\begin{figure}
\begin{centering}
\subfloat[]{\begin{centering}
\includegraphics[width=0.9\columnwidth]{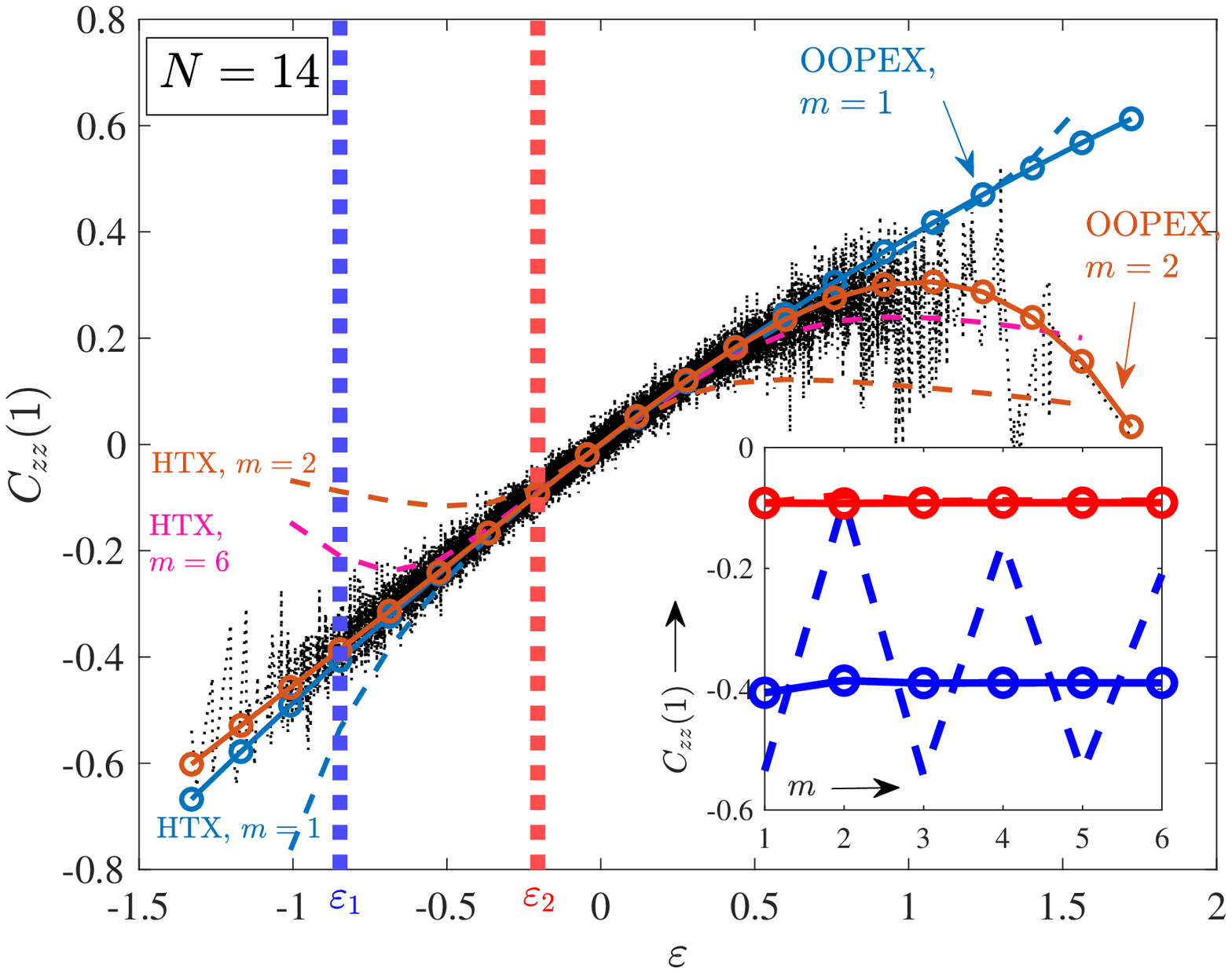}
\par\end{centering}
}
\par\end{centering}
\begin{centering}
\subfloat[]{\begin{centering}
\includegraphics[width=0.9\columnwidth]{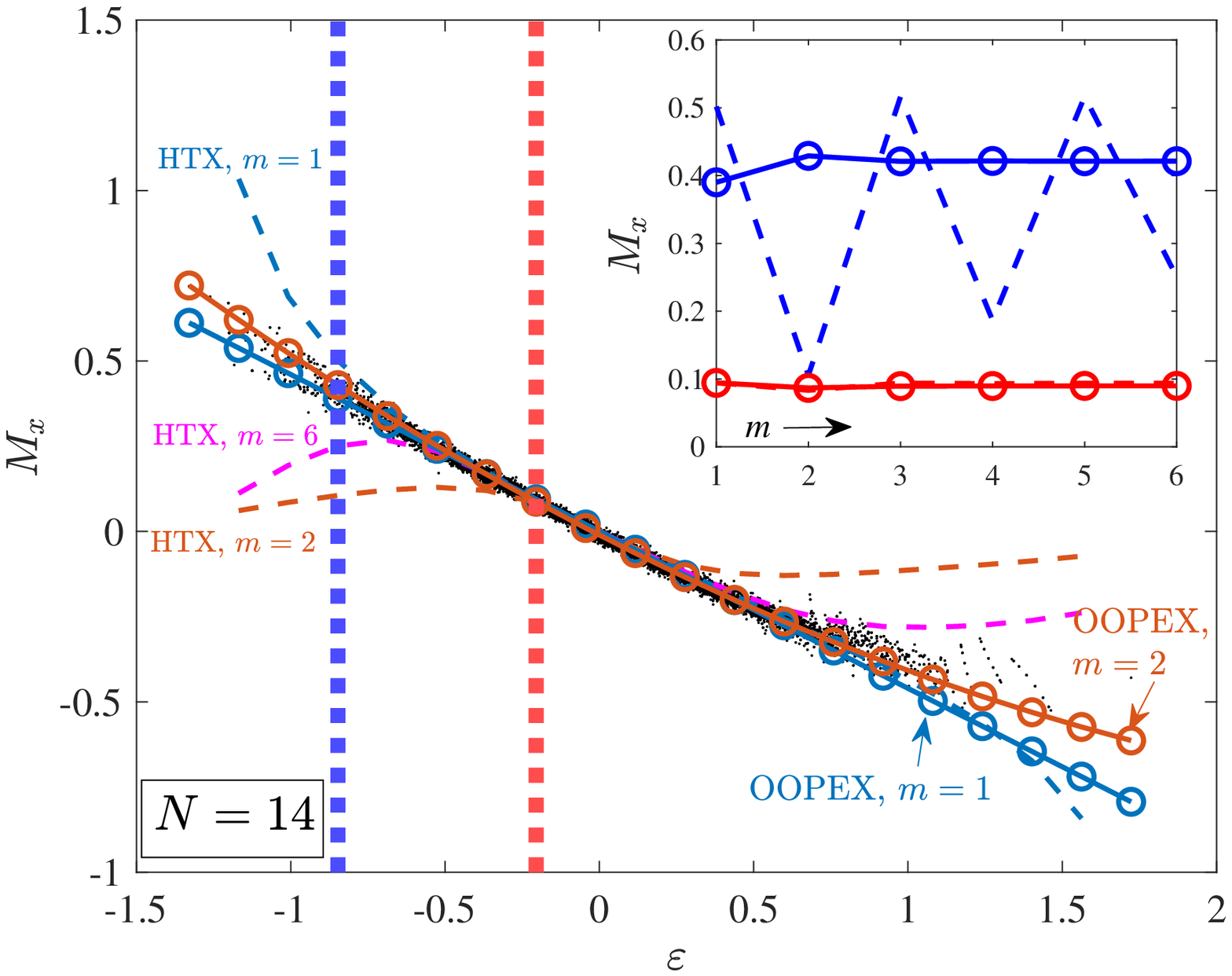}
\par\end{centering}
}
\par\end{centering}
\begin{centering}
\subfloat[]{\begin{centering}
\includegraphics[width=0.9\columnwidth]{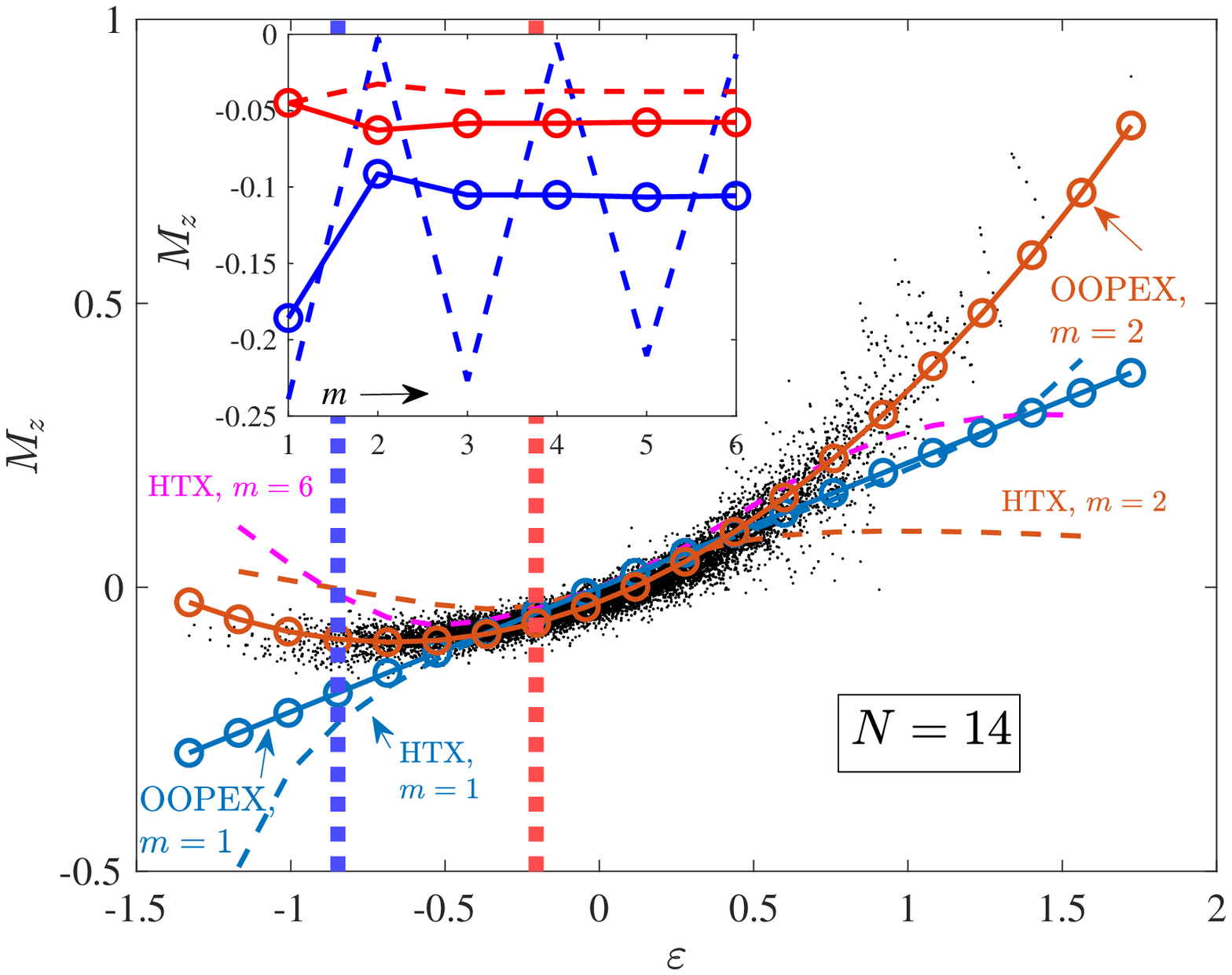}
\par\end{centering}
}
\par\end{centering}
\caption{Comparison between computations of (a) $C_{zz}(1)$, (b) $M_{x}$
and (c) $M_{z}$ using OOPEX (solid with markers), ED (dotted) and
HTX (dashed) for a $N=14$-site chain. OOPEX results for truncation
at $m=2$ agree excellently with ED whereas HTX shows significantly
worse agreement even at $m=6$. Insets show the truncation order-dependence
of the OOPEX and HTX at $\varepsilon_{1}=-0.8475$ (blue) and $\varepsilon_{2}=-0.2048$
(red) marked in the main figure. The OOPEX converges almost immediately
while HTX shows drastically poor convergence away from the middle
of the spectrum. \label{fig:Exp_values}}
\end{figure}

We study three simple representative observables: the 2-point function
$C_{zz}(\Delta r)=\frac{1}{2(N-\Delta r)}\sum_{r,r'=r\pm\Delta r}\left(\left\langle \sigma_{r}^{z}\sigma_{r'}^{z}\right\rangle -\left\langle \sigma_{r}^{z}\right\rangle \left\langle \sigma_{r'}^{z}\right\rangle \right)$
and the 1-point functions $M_{i}=\frac{1}{N}\sum_{r}\left\langle \sigma_{r}^{i}\right\rangle ,i=x,z$.
Fig. \ref{fig:Exp_values} compares the expectation values of $C_{zz}(1)$,
$M_{x}$ and $M_{z}$ using the OOPEX, ED and HTX. The match between
ED and OOPEX is striking for just $m=2$ over a wide range of energy
densities $\varepsilon=E/N$, whereas HTX deviates significantly from
ED even for $m=6$ for $|\varepsilon|\gtrsim0.5$. The inset shows
rapid convergence of the OOPEX with $m$ at both $\varepsilon_{1}=-0.8475$
and $\varepsilon_{2}=-0.2048$. In contrast, HTX converges poorly
(well) for the former (latter) $\varepsilon$; note $|\varepsilon_{1}|>0.5>|\varepsilon_{2}|$.
Since the OOPEX and HTX are both power series-based algorithms that
work best near $\varepsilon=0$, better performance of the former
is likely due to its ability to exploit the non-integrability of the
system.

\begin{figure}
\subfloat[]{\includegraphics[width=0.9\columnwidth]{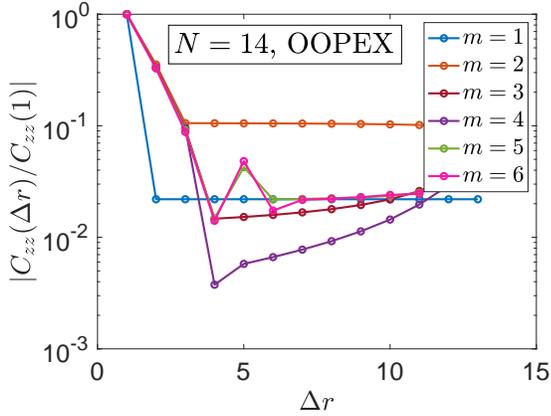}

}

\subfloat[]{\includegraphics[width=0.9\columnwidth]{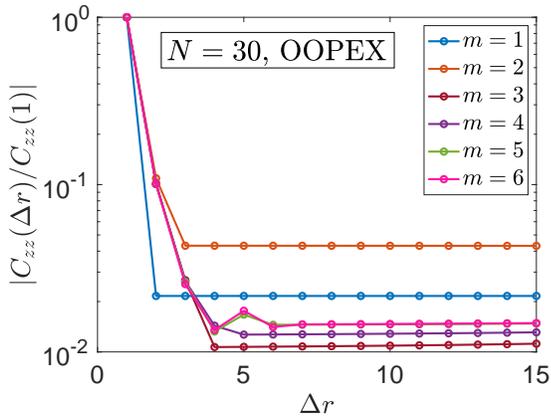}

}

\subfloat[]{\includegraphics[width=0.9\columnwidth]{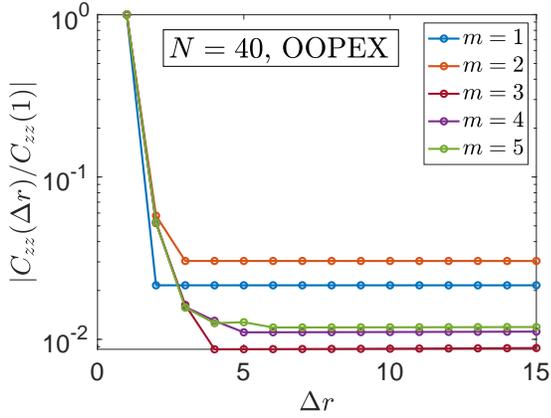}

}

\caption{Normalized $|C_{zz}(\Delta r)|$ using OOPEX at $\varepsilon_{2}$,
where HTX agrees with ED for $C_{zz}(1)$. (a) At $N=14$, strong
finite-size effects produce an upturn for $\Delta r\gtrsim4$ that
survives up to $m=6$. (b) For $N=30$, the upturn occurs near the
opposite edge, so is missing in the data shown. Moreover, $m=5,6$
data nearly overlap, indicating convergence. Finally, clear exponential
decay for $\Delta r\lesssim5$ allow extracting $\xi$ while a constant
decay rate for $m=3\dots6$ imply that $\xi$ converges for $m=3$.
(c) $N=40$ data are similar to data in (b), but the near-overlap
of $m=4,5$ data indicate faster convergence. \label{fig:Czz(delta r)}}
\end{figure}

\begin{figure}
\includegraphics[width=0.9\columnwidth]{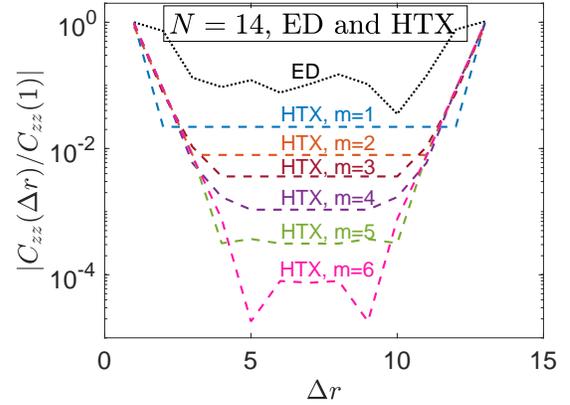}

\caption{Normalized $|C_{zz}(\Delta r)|$ at $\varepsilon_{2}$ for $N=14$
obtained using ED and HTX. ED data show strong finite-size effects
at $N=14$, making it impossible to determine $\xi$. HTX data show
exponential decay for $\Delta r\lesssim5$, but the accessible $N(\protect\leq14$)
is too small to perform finite-size scaling of $\xi$.\label{fig:Czz(delta r)-ED-HTX}}

\end{figure}

\begin{figure}
\includegraphics[width=1\columnwidth]{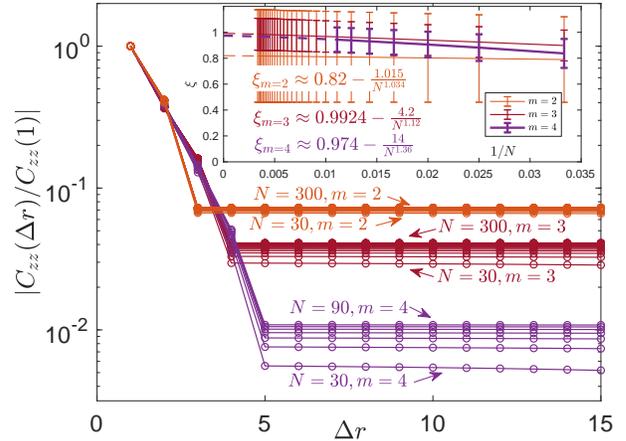}

\caption{Normalized $|C_{zz}(\Delta r)|$ at $\varepsilon_{1}$ computed using
the OOPEX for $N=30,40,\dots,300$ at $m=2,3$ and for $N=30,40,\dots,90$
at $m=4$. Exponential decay is discernible in each data set for $1\protect\leq\Delta r\protect\leq m+1$,
allowing determination of $\xi$. Inset shows finite-size scaling
of $\xi$ at various fixed $m$. For $N\to\infty$, the $m=3,4$ data
produce nearly identical values of $\xi$, indicating convergence.\label{fig:xi}}
\end{figure}

In Fig. \ref{fig:Czz(delta r)}, we examine the behavior of $C_{zz}(\Delta r)$
at $\varepsilon_{2}$, where both HTX and OOPEX concur with ED for
$C_{zz}(1)$. Figs. \ref{fig:Czz(delta r)}(a-c) show that access
to large $N$ with the OOPEX helps avoid finite-size effects and enables
extracting a correlation length $\xi$. In contrast, Fig. \ref{fig:Czz(delta r)-ED-HTX}
shows that extracting a correlation is unreliable using the HTX and
impossible using ED at $N=14$. At a lower energy density (relative
to the ground state) $\varepsilon_{1}$, HTX behaves poorly even for
$C_{zz}(1)$, as Fig. \ref{fig:Exp_values} shows. In stark contrast,
we find that the OOPEX not only works well for $C_{zz}(1)$, it works
well-enough for $C_{zz}(\Delta r>1)$ to determine $\xi$. Thus, we
extract $\xi$ at $\varepsilon_{1}$ in Fig. \ref{fig:xi} and show
that merely $m=3$ yields $\xi$ that is well-behaved in the thermodynamic
limit. The curves flatten for large $\Delta r$ because, for range-$R$
Hamiltonians, the OOPEX can compute bare 2-point correlations between
sites separated by up to $O(mR)$ sites. For larger separations, connected
correlations receive contributions only from the disconnected parts.
In the current example, this means $\left\langle \sigma_{r}^{z}\sigma_{r+\Delta r}^{z}\right\rangle =0$
for large enough $\Delta r$ so that $\left\langle \sigma_{r}^{z}\sigma_{r+\Delta r}^{z}\right\rangle -\left\langle \sigma_{r}^{z}\right\rangle \left\langle \sigma_{r+\Delta r}^{z}\right\rangle =-\left\langle \sigma_{r}^{z}\right\rangle \left\langle \sigma_{r+\Delta r}^{z}\right\rangle \approx\left\langle \sigma_{r}^{z}\right\rangle ^{2}$
upto boundary effects.

\subsection{Entanglement entropy}

\begin{figure}
\begin{centering}
\includegraphics[width=0.9\columnwidth]{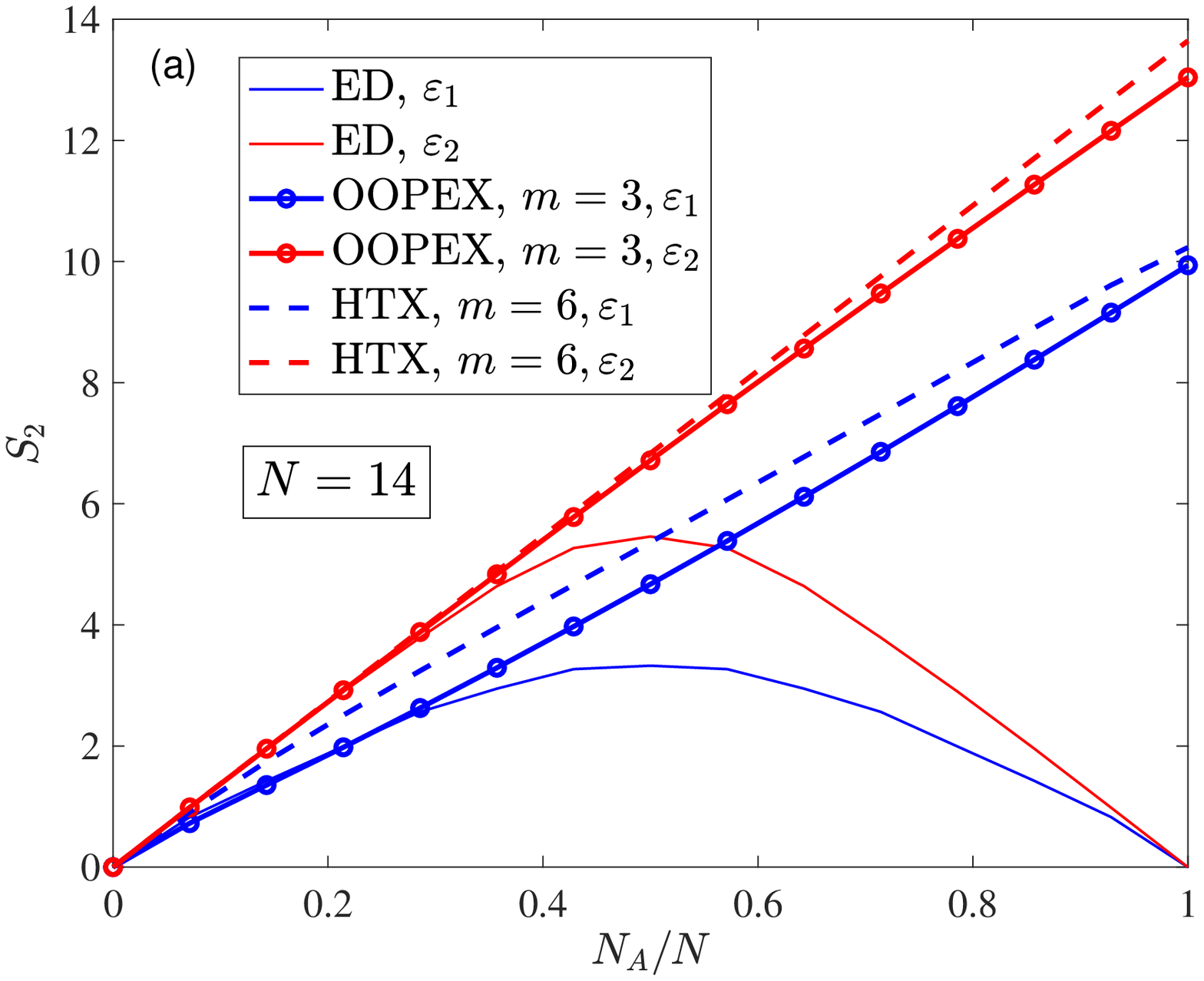}
\par\end{centering}
\caption{$S_{2}$ computed using the OOPEX, ED and HTX at $\varepsilon_{1,2}$
at $N=14$. The OOPEX matches ED better than HTX does at a smaller
truncation order.\label{fig:S2}}

\end{figure}

We now consider the second Renyi entanglement entropy $S_{2}(A)$
between the leftmost $N_{A}$ sites and the rest of the system. For
chaotic eigenstates, it is well-known that $S_{2}(A)$ follows a volume
law: $S_{2}/N_{A}=O(1)$. Fig. \ref{fig:S2} shows $S_{2}$ computed
using ED, OOPEX and HTX at $\varepsilon_{1,2}$ and $N=14$. The OOPEX
at $m=3$ shows better match with ED than HTX at $m=6$, and the linear
growth with $N_{A}$ is apparent. Since the OOPEX can access only
short-distance correlations and $S_{2}(A)=\sum_{\mathcal{O}_{\ell}\in A}\left|\text{tr}(\rho\mathcal{O}_{\ell})\right|^{2}$,
where the sum runs over operators that have support strictly in $A$,
the agreement between OOPEX and ED indicates that most of the Renyi
entropy is carried by short-distance correlations.

\begin{figure}
\begin{centering}
\subfloat[]{\begin{centering}
\includegraphics[width=0.9\columnwidth]{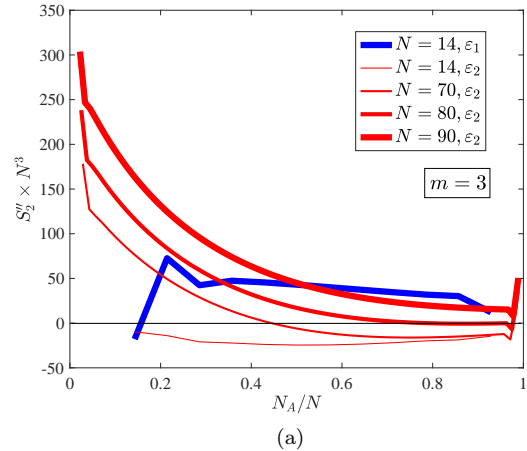}
\par\end{centering}
}
\par\end{centering}
\begin{centering}
\subfloat[]{\begin{centering}
\includegraphics[width=0.9\columnwidth]{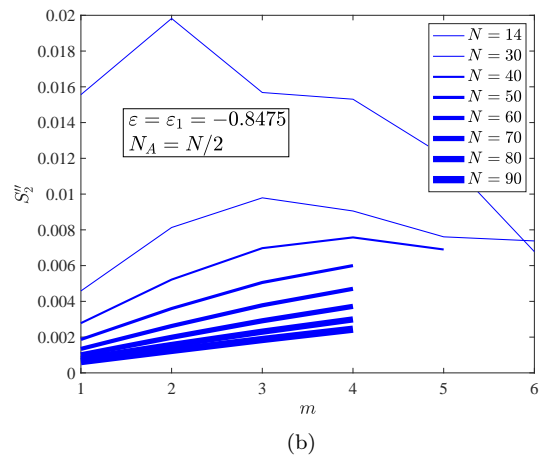}
\par\end{centering}
}
\par\end{centering}
\begin{centering}
\subfloat[]{\begin{centering}
\includegraphics[width=0.9\columnwidth]{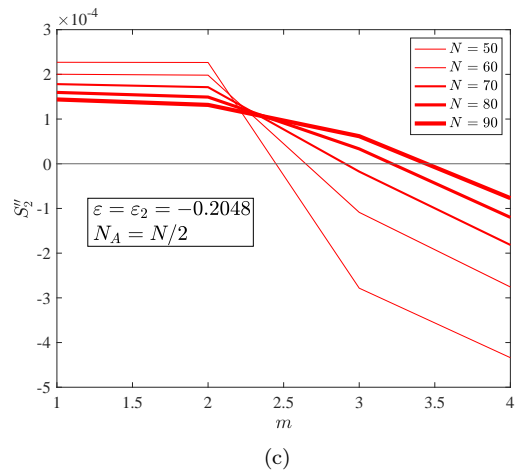}
\par\end{centering}
}
\par\end{centering}
\centering{}\caption{$S_{2}^{\prime\prime}$ computations using the OOPEX. (a) $N^{3}S_{2}^{\prime\prime}$
vs $N_{A}/N$ for $N=14,\varepsilon=\varepsilon_{1}$ and various
$N$ at $\varepsilon=\varepsilon_{2}$. For given $(m,\varepsilon)$,
accessing a large enough $N$ gives a state with $S_{2}^{\prime\prime}>0$.
For $(m,\varepsilon)=(3,\varepsilon_{1})$, $N=14$ suffices for almost
all $N_{A}$, while $(m,\varepsilon)=(3,\varepsilon_{2})$ requires
$N\gtrsim80$. The $N^{3}$ factor ensures clarity of the plot. (b)
$m$-dependence of $S_{2}^{\prime\prime}$ at $\varepsilon=\varepsilon_{1}$,
$N_{A}=N/2$ for several $N$. Although data does not converge, $S_{2}^{\prime\prime}>0$
always and increases with $m$ at larger $N$. (c) $m$-dependence
of $S_{2}^{\prime\prime}$ at $\varepsilon=\varepsilon_{2}$, $N_{A}=N/2$
for several $N$. Increasing $m$ turns $S_{2}^{\prime\prime}$ negative,
but increasing $N$ at fixed $m$ tends the data towards a positive
$S_{2}^{\prime\prime}$.\label{fig:S2''}}
\end{figure}

Recent work argued that $S_{2}^{\prime\prime}=d^{2}S_{2}/dN_{A}^{2}>0$
for chaotic eigenstates \citet{LuGrover2019}. Unfortunately, the
hard constraints $S_{2}(0)=S_{2}(N)=0$ and the positivity of $S_{2}$
force $S_{2}^{\prime\prime}<0$ when computed using ED at the small
$N$ it can access. However, the large $N$ accessible with the OOPEX
and the fact that the $\rho$ produced by the OOPEX remove the constraint
$S_{2}(N)=0$ and enable observing $S_{2}^{\prime\prime}>0$. As shown
in Fig. \ref{fig:S2''}(a), $S_{2}^{\prime\prime}>0$ for almost all
$N_{A}$ already at $N=14$ when $\varepsilon=\varepsilon_{1}$. When
$\varepsilon=\varepsilon_{2}$, $S_{2}^{\prime\prime}<0$ for small
$N$, but becomes $>0$ when $N\gtrsim80$. Thus, access to a large
$N$ with the OOPEX is key for detecting the convexity of $S_{2}$. 

The positivity of $S_{2}^{\prime\prime}$ obtained using the OOPEX,
however, must be taken with a grain of salt. Fig. \ref{fig:S2''}(b)
shows the $m$-dependence of $S_{2}^{\prime\prime}$ for several values
of $N$ at $\varepsilon=\varepsilon_{1}$. Although $S_{2}^{\prime\prime}>0$
for all the cases shown, the data clearly have not converged. However,
$S_{2}^{\prime\prime}$ grows with $m$ at larger $N$, suggesting
that $S_{2}^{\prime\prime}$ will probably converge to a positive
value. The behavior is less clear at $\varepsilon=\varepsilon_{2}$,
as shown in Fig. \ref{fig:S2''}(c). Now, $S_{2}^{\prime\prime}$
decreases with $m$ and becomes negative for $m=4$ for all accessible
$N$, while increasing $N$ at fixed $m$ increases $S_{2}^{\prime\prime}$.
It is, thus, plausible that $S_{2}^{\prime\prime}>0$ once convergent
results have been obtained in the thermodynamic limit, but our program
is currently unable settle this issue.

\subsection{Computational cost}

\begin{figure}
\begin{centering}
\subfloat[]{\begin{centering}
\includegraphics[width=0.9\columnwidth]{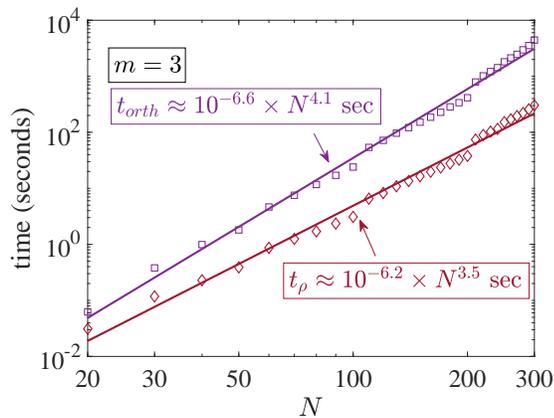}
\par\end{centering}
} 
\par\end{centering}
\begin{centering}
\subfloat[]{\begin{centering}
\includegraphics[width=0.9\columnwidth]{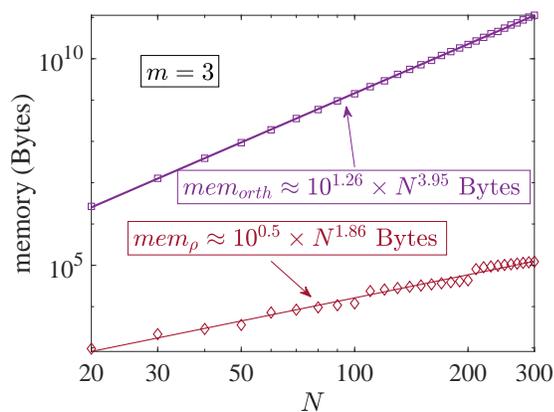}
\par\end{centering}
} 
\par\end{centering}
\caption{Scaling of time (a) and memory (b) requirements with $N$ at fixed
$m=3$. $t_{orth}$ ($t_{\rho}$) denotes the time required to compute
the orthonormalized Krylov space (compute $\rho(E)$ at fixed $E$
given the orthonormalized Krylov space), while $mem_{orth}$ ($mem_{\rho}$)
denotes the minimum memory needed to compute this space (to store
$\rho(E)$ at fixed $E$).{\small{}\label{fig:Scaling}}}
\end{figure}

Fig. \ref{fig:Scaling} shows that time and memory needs of the OOPEX
for $m=3$ scale as power laws in $N$ with modest exponents. In particular,
the time and memory needed to create the orthonormalized Krylov space
(to compute $\rho(E)$ for a fixed $E$ given the orthonormalized
Krylov space) grow as $t_{orth}\sim N^{4.1}$ ($t_{\rho}\sim N^{3.5}$)
and $mem_{orth}\sim N^{3.95}$ ($mem_{\rho}\sim N^{1.86}$). Computing
$\left\langle A(E)\right\rangle $ given $\rho(E)$ is practically
instantaneous. However, multiplying $H$ with itself $m-1$ times
to create the Krylov space demands resources that grow exponentially
with $m$, which limits computations to relatively small $m$. 

\section{Conclusion}

In conclusion, we have introduced an algorithm, the OOPEX, that can
compute expectation values in chaotic eigenstates with polynomial
effort, and demonstrated it on a prototypical model. The algorithm
converges rapidly thanks to the ETH, and gives access to system sizes
of several hundred sites, thus enabling computations of correlation
lengths that were beyond the capabilities of ED. Detailed comparisons
with other algorithms including quantum Monte Carlo methods \citep{Troyer2005,SignProblem},
finite temperature density matrix renormalization group \citet{Feiguin2005,Jiang:2020aa,Karrasch_2013,Jansen2020}
and the kernel polynomial method \citet{WeissKPMReview} will be presented
in future work. 

The OOPEX should be most useful for investigating physics in interacting
regimes where $\xi$ is finite and the $E$-dependence of physical
quantities is smooth, such as finite temperature physics above quantum
critical points and theories with a holographic gravitational dual.
The fundamental reliance of the OOPEX on the ETH implies that it could
be also a useful sensor of ergodicity breaking and, for example, effectively
probe the many-body localization transition from the chaotic side.
Finally, while extracting critical exponents associated with phase
transitions may be challenging for the OOPEX, it might help identify
the \emph{presence} of a phase transition via a broad peak in the
$E$-dependent correlations of the order parameter. These problems
will be investigated in the future.

\begin{acknowledgments}
We acknowledge invaluable discussions with Xiao-Liang Qi, Ashvin Vishwanath,
Scott Aaronson and especially Hitesh Changlani and Fabien Alet. We
acknowledge support from the Division of Research, Department of Physics
and the College of Natural Sciences and Mathematics at the University
of Houston and from NSF-DMR-2047193. 
\end{acknowledgments}

 \bibliographystyle{apsrev4-1}
\bibliography{OOPEX_refs}

\end{document}